\documentclass[fleqn,usenatbib]{mnras}

\usepackage{newtxtext,newtxmath}

\usepackage[T1]{fontenc}

\DeclareRobustCommand{\VAN}[3]{#2}
\let\VANthebibliography\thebibliography
\def\thebibliography{\DeclareRobustCommand{\VAN}[3]{##3}\VANthebibliography}

\usepackage{graphicx}	
\usepackage{amsmath}	
\defcitealias{Eltvedt2023}{Paper~I}
\defcitealias{EltvedtPaper2}{Paper~II}

\title[VST ATLAS QSO SURVEY III]{The VST ATLAS Quasar Survey III: Halo mass function via quasar clustering  and quasar-CMB lensing cross-clustering}

\author[A. M. Eltvedt et al.]
{Alice M. Eltvedt,$^{1}$\thanks{E-mail: aeltvedt@alumni.princeton.edu}
T. Shanks,$^{1}$\thanks{E-mail: tom.shanks@durham.ac.uk}
N. Metcalfe,$^{1}$
B. Ansarinejad,$^{2}$
L.F. Barrientos,$^{3}$\newauthor
D.N.A. Murphy$^{3,4}$
and D.M. Alexander$^{1}$
\\
$^{1}$Centre for Extragalactic Astronomy, Department of Physics, Durham University, South Road, Durham, DH1 3LE, UK\\
$^{2}$ School of Physics, University of Melbourne, Parkville, VIC 3010, Australia\\
$^{3}$Instituto de Astrofisica, Facultad de Fisica, Pontificia Universidad Catolica de Chile, Santiago, Chile\\
$^{4}$Institute of Astronomy, University of Cambridge, Madingley Road, Cambridge CB3 0HA, UK \\
}

\date{Accepted XXX. Received YYY; in original form ZZZ}

\pubyear{2024}

\begin{document}
\label{firstpage}
\pagerange{\pageref{firstpage}--\pageref{lastpage}}
\maketitle

\begin{abstract}
We exploit the VST ATLAS quasar/QSO catalogue to perform three 
measurements of the quasar halo mass profile. First, we make a new
estimate of the angular auto-correlation function of $\approx230,000$
ATLAS quasars with $z_{photo}\lesssim 2.5$ and $17<g<22$. By comparing
with the $\Lambda$CDM  mass clustering correlation function, we measure
the quasar bias to be $b_Q\approx2.1$, implying a quasar halo mass of
$M_{halo}\approx8.5\times10^{11}$h$^{-1} M_\odot$. Second, we
cross-correlate these $z\approx1.7$ ATLAS quasars with  the Planck
Cosmic Microwave Background (CMB) lensing maps, detecting a somewhat
stronger signal at $4'<\theta<60'$ than previous authors. Scaling these authors' model fit  to our data we estimate a quasar host halo mass of
$M_{halo}\approx8.3\times10^{11}h^{-1}$M$_{\odot}$.  Third, we fit  Halo
Occupation Distribution (HOD) model  parameters to  our quasar
auto-correlation function and from the derived halo mass function
we estimate a quasar  halo mass of
$M_{halo}\approx2.5\times10^{12}$h$^{-1} M_\odot$. We then compare our 
HOD model prediction to our quasar-CMB lensing result, confirming
their consistency. We  find that most ($\approx2/3$) QSOs have halo masses
within a factor of $\approx3$ of this average mass. An analysis based on
the probability of X-ray detections of AGN in galaxies and the galaxy
stellar mass function gives a similarly small mass range. Finally, we
compare the quasar halo mass and luminosity  functions and suggest that 
gravitational growth may produce the constant space density with 
redshift seen in the quasar luminosity function.
\end{abstract}




\section{Introduction}

Quasars are great tracers of large scale structure  as they are visible across a wide redshift range and their high bias also makes  their clustering easier to detect. In \cite{EltvedtPaper2} (hereafter \citetalias{EltvedtPaper2})  we also exploited weak lensing `magnification bias' of quasars to measure the mass profiles of galaxies and clusters of galaxies. But in terms of measuring the masses and halo profiles of quasars themselves, we must turn to the Cosmic Microwave Background (CMB) as the CMB provides a background to all of the observable structures in the universe, including quasars. We can use the deflection of CMB photons to trace  the matter density field (e.g. \citealt{BlanchardSchneider1987,Seljak1996}) which, when combined with the QSO clustering measurements, allows the quasar bias to be calculated. We can then evaluate the relationship between this bias and the host dark matter halo mass and place the quasars in a cosmological context. Here we follow the recent works of \cite{Geach2019, Han2019, Petter2022, Petter2023}, who argue that the cross-correlation of a  CMB lensing convergence map and a quasar sample offers a more direct way of estimating quasar bias than the more commonly used way of combining standard cosmology assumptions with measurements of the quasar 2-point auto-correlation function (e.g. \citealt{Croom2005, Chehade2016}), with a lower level of systematics. So here we aim to measure the halo mass profiles of QSOs from the catalogue of \citetalias{Eltvedt2023}. These were first used to measure the halo mass profiles of galaxy clusters, galaxies and LRGs via QSO lensing in \citetalias{EltvedtPaper2} and now these QSOs themselves will have their halo mass profiles measured by fitting HODs to the QSO clustering correlation function and then testing this  fit using QSO-CMB lensing.
 
The overall aim therefore is to determine the QSO host halo mass function at $z\approx1.7$. Previously, the main route to finding QSO halo masses has been by estimating the bias of QSOs through the comparison of the QSO 3-D redshift-space correlation function with an assumed $\Lambda$CDM clustering correlation function, and then using the bias-halo mass relation to estimate average halo masses. This then leads to studies investigating, for example, how QSO halo mass may depend on QSO luminosity \citep{Chehade2016}. 

Here, we first measure the 2-D angular auto-correlation function of our QSO sample defined in \citetalias{EltvedtPaper2}. This function is independent of redshift space distortions and only one or two authors have previously been able to go this route (e.g. \citealt{Petter2023}). The high quality of the VST ATLAS QSO samples have allowed us to measure the QSO 2-D angular correlation function and make a new bias and average halo mass estimates in this way. But this route only delivers the average halo mass at a given redshift with no information about e.g. the width of the mass distribution. We therefore proceed to make a more detailed estimate of the QSO halo mass function by fitting QSO HOD models to the QSO auto-correlation function data.

We then go on to use QSO lensing of the CMB to make further tests of the above results. Here, we cross-correlate the ATLAS QSO sample with the {\it Planck} CMB lensing convergence map. This first allows us to test directly the bias estimated from QSO clustering, free from the assumption of the $\Lambda$CDM model, and to make an independent estimate of the QSO average host halo mass via the above bias-halo mass relation. Then, we make a direct measurement of the full QSO host halo mass function by fitting a HOD model to our QSO-Planck cross-correlations.  Finally, we are able to determine the QSO halo mass function by multiplying the derived QSO HOD with the $\Lambda$CDM halo mass function.

The structure of this paper is as follows. In Section~\ref{sec:data} we describe the two data catalogues we will be using throughout the paper. In Section~\ref{sec:wqq} we start our analysis by performing an auto-correlation of our QSO sample to measure the clustering amplitude and compare our sample to that of \cite{Petter2023}, 
before deriving both the QSO bias and host halo mass via this $\omega_{qq}$. We then detect the deflection (lensing) of the Cosmic Microwave Background (CMB) \citep{Planck2018} by our quasar candidates through a cross-correlation of our QSO survey and the Planck CMB lensing map (e.g. \citealt{Geach2019,Han2019, Petter2022}) in Section~\ref{sec:G19cmb} to also derive a host halo mass via the measurement of the QSO bias from a simple scaling of the halo profile model of \cite{Geach2019}. Finally, in Section~\ref{sec:HOD_model_intro} we fit both our $\omega_{qq}$ and $\omega_{Q-CMB}$ results with a HOD model to find the best fitting halo mass distribution of our QSO sample. We present our conclusions in Section~\ref{sec:conclusions}. Throughout, we assume a standard, spatially flat, cosmology with $\Omega_m=0.3$ and  a Hubble constant assumed to be 100 h km s$^{-1}$ Mpc$^{-1}$, with h=0.7 unless otherwise stated.\\


\section{Data}
\label{sec:data}

\subsection{Quasar Sample}
\label{sec:xcorr_QSO_sample}

\begin{figure}
    \centering
    \includegraphics[width=\columnwidth]{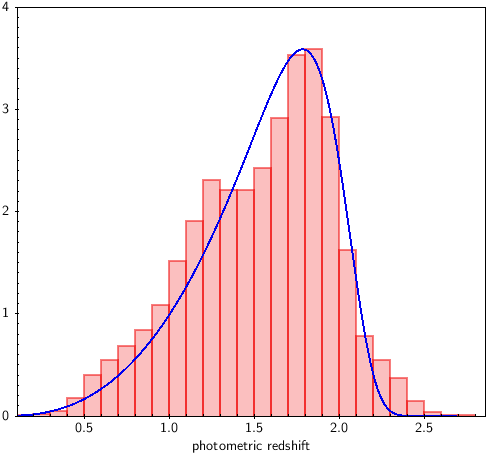} 
	\caption{Our VST ATLAS+unWISE QSO sample redshift distribution,  $dn(z)/dz$, along with the SDSS $dn(z)/dz$ redshift distribution from \protect\cite{Scranton2005} described by Equation~\ref{eq:redshift_distribution}, with the y-axis showing the total number of QSOs in the NGC+SGC divided by 7049.4 in order to scale it to the model.}
	\label{fig:QSO_redshift}
\end{figure}

The VST-ATLAS quasar catalogue described in \citetalias{Eltvedt2023} has a certain amount of stellar and galaxy contamination to improve completeness. For the purposes of the analyses in this paper, we use a non-photometric redshift restricted version of the QSO sample described in \citetalias{EltvedtPaper2}, which introduces restricted $ugriW1W2$ selections to the Priority 1 QSO sample from \citetalias{Eltvedt2023}, as well as a mask to remove Tycho stars and globular clusters.
These selections result in a total of $230914$ ATLAS quasar candidates giving us a sky density of ~$49$deg$^{-2}$. The photometric redshift distribution of our final QSO sample is shown in Fig. ~\ref{fig:QSO_redshift}, showing a clear peak in the distribution at $\approx$1.8. Also shown in Fig.~\ref{fig:QSO_redshift}, is the  $dn(z)/dz$  distribution for SDSS quasars to a similar limit  in the form of:

\begin{equation}
      \left.\frac{dn}{dz}\right\vert_Q\sim z^{2.56}\exp[-(\frac{z}{2.02})^{12.76}], 
     \label{eq:redshift_distribution}
\end{equation} \\

\noindent \citep{Scranton2005} which, although the model is formally rejected by the data, it still represents a reasonable fit to the shape of our distribution. We therefore use this relation to describe our QSO sample in  Section~\ref{sec:HOD_model_intro}.

\begin{figure}
    \centering
 	\includegraphics[width=\columnwidth]{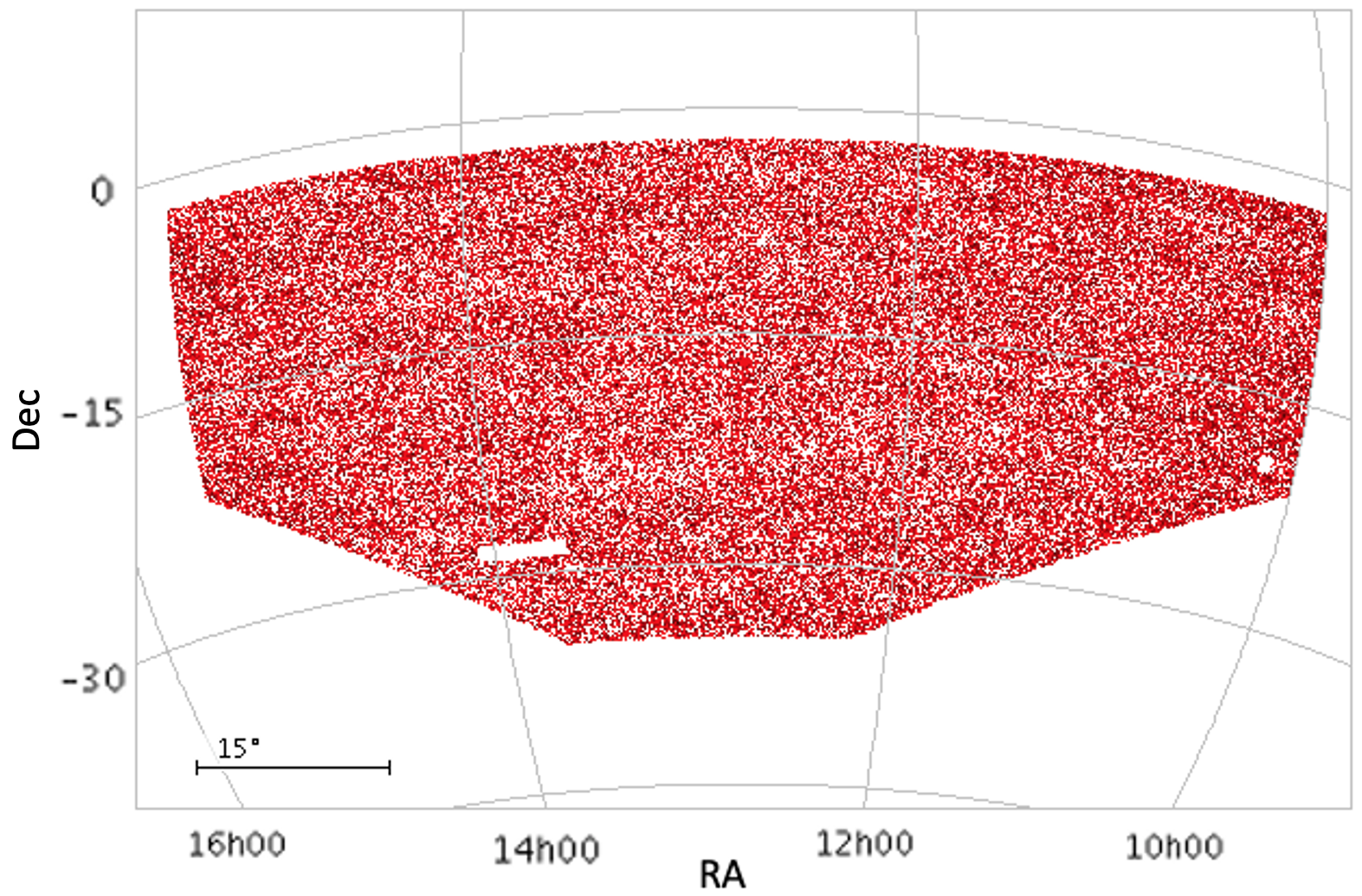} 
    \includegraphics[width=\columnwidth]{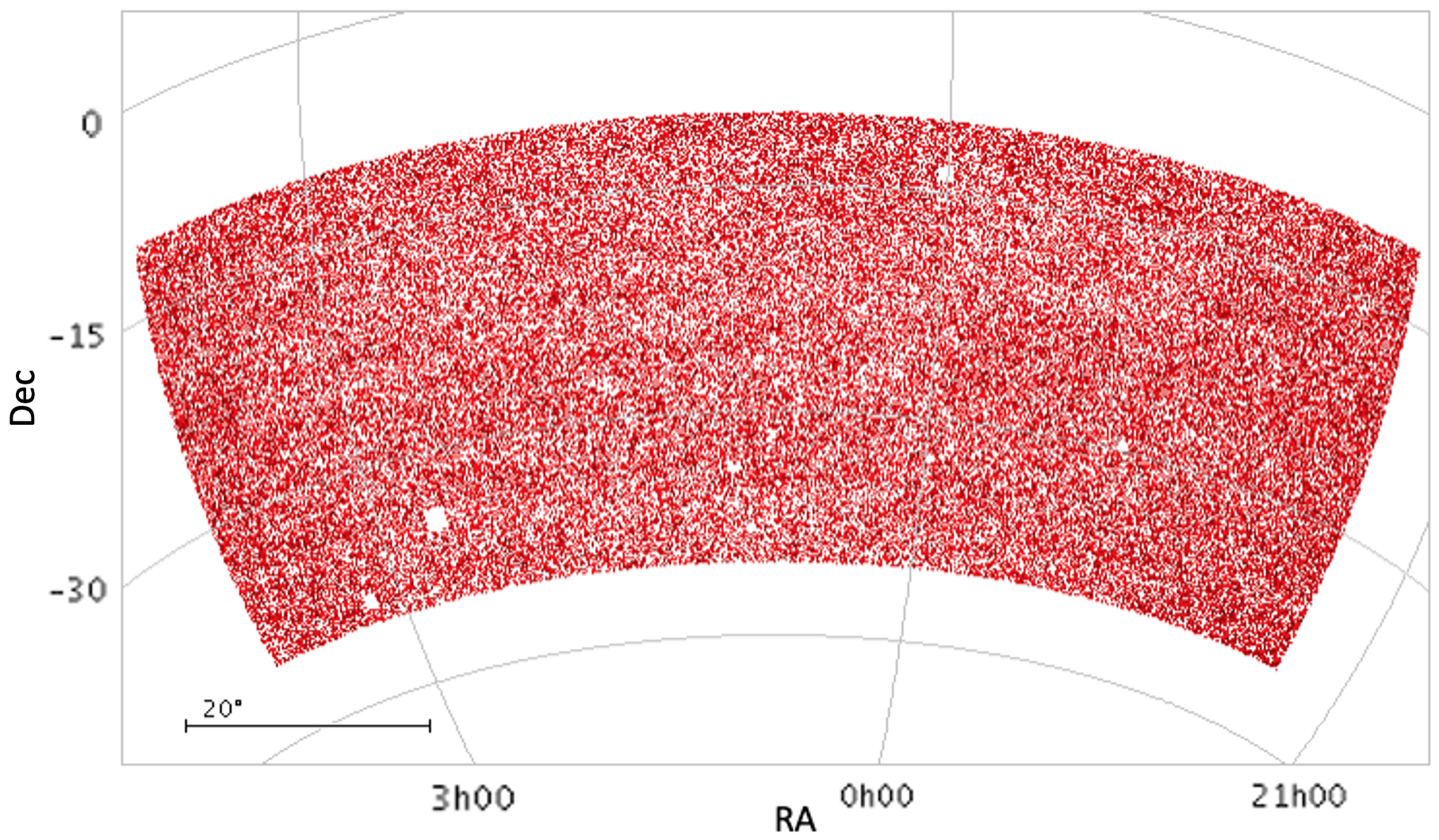} 
 	\caption[Sky maps of the VST ATLAS QSO candidates in the NGC (above) and the SGC (below) used in this paper. Areas of higher density are shaded in a darker red whereas lower density areas are lighter.]{Sky density maps of the VST ATLAS QSO candidates in the NGC (above) and the SGC (below) used in this paper. Areas of higher density are shaded in a darker red whereas lower density areas are lighter.}
	\label{fig:NGCSGC_qsomap.png}
\end{figure}

We visually inspect the QSO candidate distribution in the sky to see a relatively flat distribution of candidates across the sky as well as confirming that we have masked out bright stars and globular clusters. This density of candidates is shown in Fig.~\ref{fig:NGCSGC_qsomap.png}. \\


\subsection{Planck CMB Lensing Convergence Map}

\begin{figure}
    \centering
 	\includegraphics[width=\columnwidth]{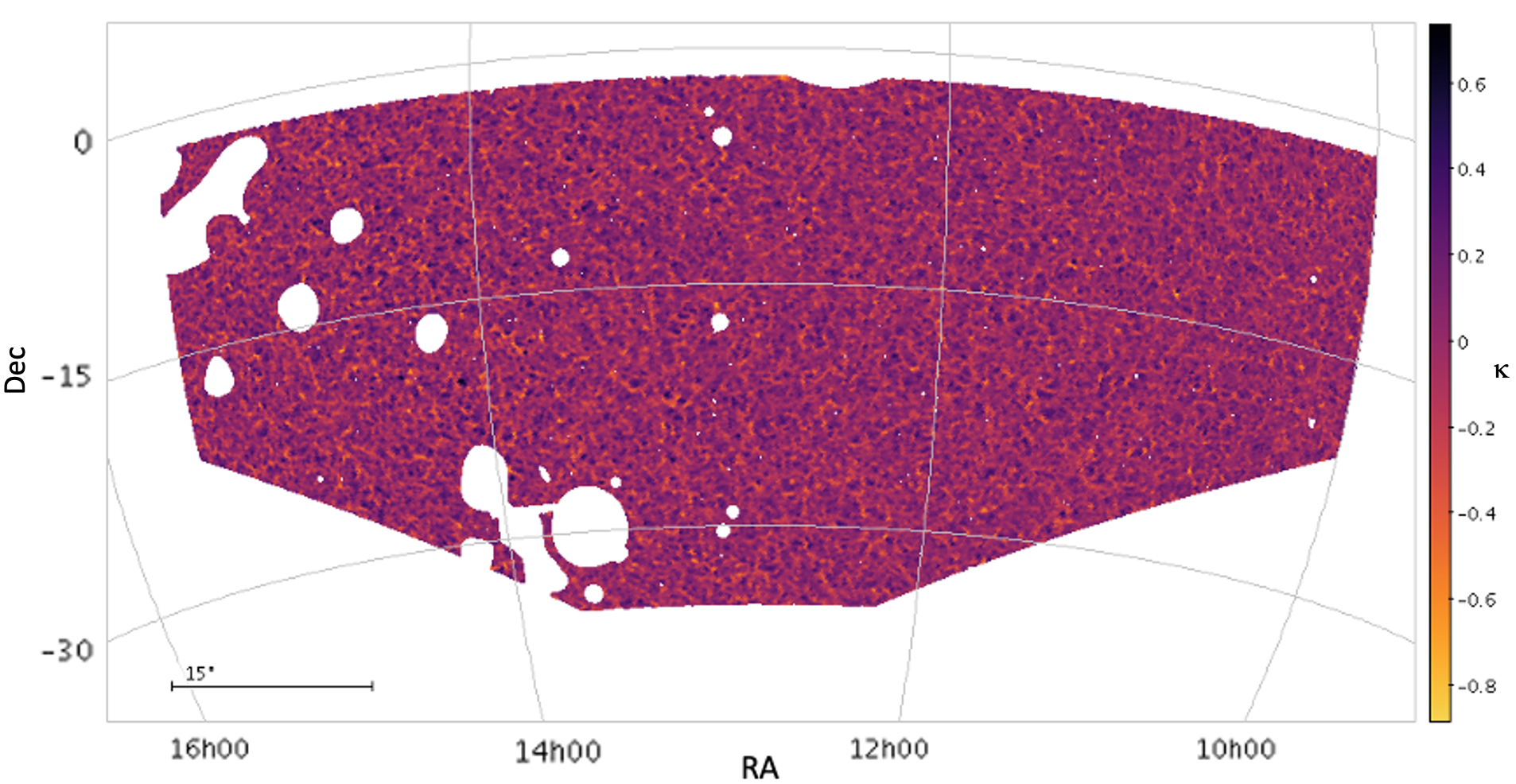}
    \includegraphics[width=\columnwidth]{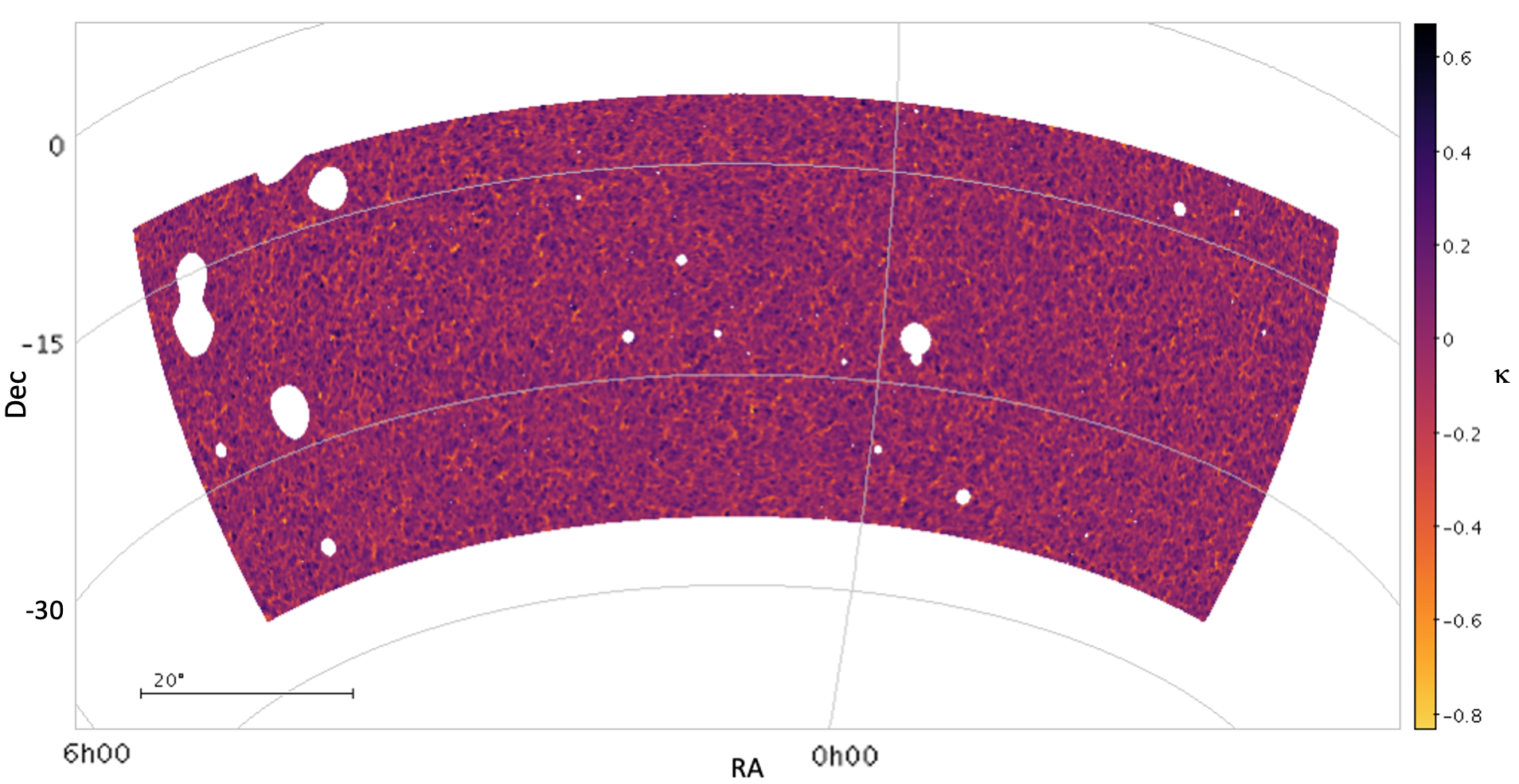}
 	\caption[The CMB lensing convergence map in the NGC and SGC.]{The CMB lensing convergence map in the NGC (above) and SGC (below).}
	\label{fig:NGCSGCcmb_lensingmap.png}
\end{figure}

\noindent To perform our cross-correlation, we use the 2018 release of the Planck lensing convergence baseline map \citep{Planck2018}, with nside$=2048$ and $l_{max}=4096$, as described in \citetalias{EltvedtPaper2}. Within the healpy {\it alm2map} routine, we first smoothed the lensing spherical harmonics, $a_{lm}$, assuming a Gaussian kernel of FWHM $15'$. After using  {\it alm2map}  to convert the CMB-only minimum variance $a_{lm}$ estimates of the lensing signal to an RA +Dec file with coordinates of the Healpix pixel centers, we apply the lensing mask provided by the \cite{Planck2018} to the CMB data and select two areas that overlap our $\sim4700$deg$^2$ QSO sample, shown in Fig.~\ref{fig:NGCSGC_qsomap.png}. The resulting lensing convergence maps we use in our work are shown in Fig.~\ref{fig:NGCSGCcmb_lensingmap.png}.  Note that here we are following the method of \cite{Geach2019} except that we do not project the convergence map onto a tangential (Zenithal Equal Area) flat sky projection, since our  aim is to apply the cross-correlation technique directly without, as an intermediate step, producing images of the convergence maps around each quasar for stacking.


\section{QSO Angular Correlation Function}
\label{sec:wqq}

The QSO angular correlation function, $w_{qq}(\theta)$, measures the strength of quasar clustering as a function of the angular separation, $\theta$,  of quasars on the sky. The large width of the QSO $n(z)$, typically  spanning the range $1\lesssim z\lesssim 2$, tends strongly to dilute the QSO clustering in 3-D, making $w_{qq}$ hard to detect against the random noise and low-level systematics. Therefore QSO clustering has traditionally only been measured using QSO redshift surveys like 2dF \citep{Croom2005} which do not suffer from projection effects. Only recently have QSO angular correlation functions been measured (e.g. \citealt{Petter2023}) since they demand large sky areas to remove statistical  noise and careful treatment of QSO survey systematics. So a significant detection of $w_{qq}$ also represents a challenging test of the reliability of a quasar survey. \\

Once detected, $w_{qq}$ offers an alternative route to the quasar spatial correlation function, $\xi(r)$, uncontaminated by redshift space  distortion effects \citep{Kaiser1987}. The first aim then is usually to compare the clustering amplitudes of the QSO and underlying matter to determine the QSO linear bias, defined by $b_Q=\sqrt(\xi_{qq}/\xi_{mm})$ where $\xi_{mm}$ is the spatial correlation function of the matter distribution. The halo mass of the QSO host galaxy can then be estimated via the QSO bias - halo mass relation. Although the quasar angular correlation function, $w_{qq}$, the projected version of $\xi_{qq}$, is only detectable at lower S/N than $\xi_{qq}$, it has the advantage of being unaffected by z-space distortions and so  provides a viable alternative route to determining quasar halo masses.

\subsection{Method}

We perform the autocorrelation of our QSO sample using the Correlation Utilities and Two-point Estimates (CUTE) code \cite{CUTE2012}. As described in \citetalias{EltvedtPaper2}, CUTE calculates the autocorrelation by using the normalized \cite{LS1993} estimator, defined as:

\begin{equation}
     \omega_{QQ}(\theta)=\frac{D_QD_Q-D_QR_Q-R_QR_Q}{R_QR_Q}, 
     \label{eq:LScorrelation}
\end{equation} \\

\noindent   Here the number of randoms $R_Q$ are always $>10\times$ larger than the number of quasars, $D_Q$, i.e. $>2309140$ for the full quasar sample. We check the output generated by the Landy-Szalay estimator by manually checking the individual outputs needed to calculate the angular cross-correlation. The standard errors of the cross-correlation are estimated by using the field-field error defined as:

\begin{equation}
     \sigma_{\Bar{\omega}(\theta)}=\frac{\sigma_{N_{s}-1}}{\sqrt{N_{s}}}=\sqrt{\frac{\sum(\omega_{i}(\theta)-\Bar{\omega_{i}}(\theta))^{2}}{N_{s}^{2}-N_{s}}}, 
     \label{eq:standard_error}
\end{equation} \\

\noindent  where $N_s=8$, made up of 4 regions of approximately equal area  taken from each of the NGC and SGC quasar sub-samples (see Fig. 1 of \citetalias{EltvedtPaper2}).

 Our QSO autocorrelation result is shown in Fig.~\ref{fig:wqq}  alongside the angular autocorrelation measurements obtained by \cite{Petter2023}, who split their sample up into unobscured and obscured QSOs. Here we see that both of the \cite{Petter2023}  samples display a steeper angular autocorrelation, especially at small $\theta<0.'5$ scales, where the 1-halo term dominates. The sample used by \cite{Petter2023} is a WISE selected QSO sample, which is then matched to the DESI DR9 $r$-band \citep{Dey2019}. This has a magnitude limit of $r\sim 24$. The $r-W2$ cut made in their sample to define the obscured and unobscured samples is shown in their Figure 1, and the redshift distributions of the resulting samples are shown in their  Figure 3 (with distributions ranging from $0\lesssim z \lesssim 3.5$). The $n(z)$ distribution of obscured QSOs can be seen to be broader with a less distinct peak than our QSO $n(z)$ distribution in Fig. \ref{fig:QSO_redshift}.

\subsection{Limber's projection formula}
\label{sec:Limber_approx}
\begin{figure}
    \centering
 	\includegraphics[width=\columnwidth]{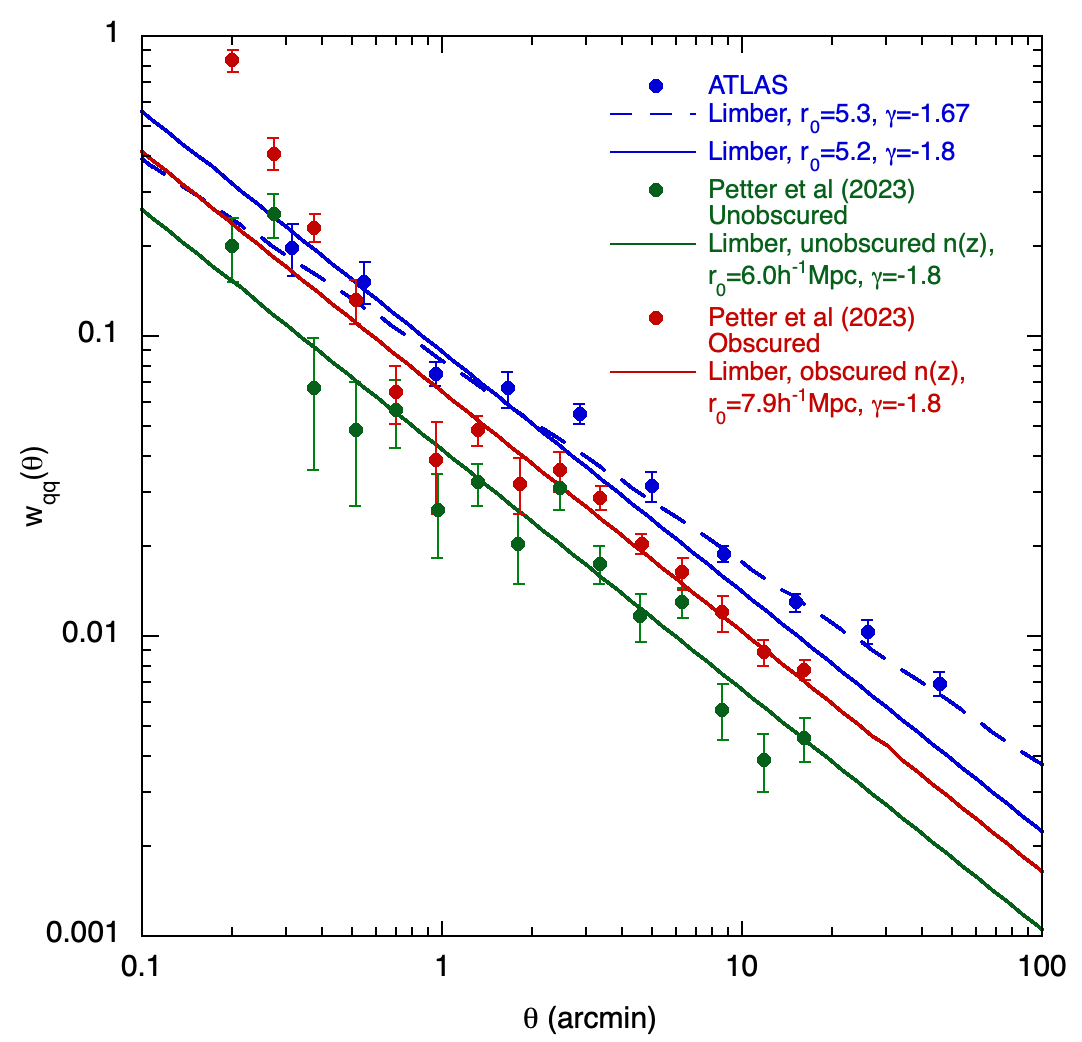} 
	\caption{We show the autocorrelation function of our ATLAS QSO sample along with the unobscured and obscured QSO samples of \protect\cite{Petter2023}. Limber's formula over full redshift range of our ATLAS QSO sample gives $r_0=5.2$ h$^{-1}$ Mpc. The Limber formula predictions of $r_0=6.0, 7.9$ h$^{-1}$Mpc for the unobscured and obscured \protect\cite{Petter2023} samples are also shown, based on their self-consistent redshift distributions. All $w_{qq}$ models assume power-laws for the 3-D $\xi(r)$ with slope $\gamma=-1.8$, along with one better fit model with $r_0=5.3$ h$^{-1}$ Mpc and slope $\gamma=-1.67$ for our QSO $w_{qq}$.}
	\label{fig:wqq}
\end{figure}

We fit all three autocorrelations via Limber's approximate projection formula \citep{Limber1953}  to translate the 2-D angular correlation function, $\omega(\theta)$ to the 3-D spatial correlation function, $\xi(r)$. If $\xi(r)$ is a power-law, $\xi(r)=(r/r_0)^{-\gamma}$, then $w(\theta)$ is also a power law of the form, $w(\theta)=(\theta/\theta_0)^{1-\gamma}$. Then, knowing the amplitude ($\theta_0$) and slope ($1-\gamma$) of the 2-D correlation function, the 3-D clustering amplitude ($r_0$) can be obtained via Limber's formula for each of the QSO samples, using the binned $n(z)$ shown in Fig. \ref{fig:QSO_redshift}. Previous studies, such as \cite{Phillipps1978} and \cite{Peebles1980}, have shown this approximation to be accurate at small angular scales below a few degrees separation if the power law is an accurate descriptor of $\xi(r)$. We see in Fig.~\ref{fig:wqq} that our QSO correlation function gives a clustering amplitude of $r_0=5.2$ h$^{-1}$ Mpc, with $\gamma =-1.8$, similar to the clustering amplitude of galaxies. The unobscured and obscured QSOs from \cite{Petter2023} give $r_0=6.0$ and $7.9$ h$^{-1}$Mpc respectively. We are able to use this clustering amplitude in the following section in order to estimate the QSO bias, and therefore derive the halo mass.


Although our QSO sample seems to have a higher angular clustering amplitude in Fig. \ref{fig:wqq}, this best-fit 3-D clustering amplitudes, $r_0$, would indicate the opposite. This occurs because the ATLAS QSO $n(z)$ has a smaller width than the $n(z)$'s of the \cite{Petter2023} QSO samples, shown in their Figure 3. If a QSO sample genuinely has a higher 3-D clustering amplitude, this would imply a higher QSO halo bias. But it could also indicate that the QSO samples of \cite{Petter2023} have lower star contamination or they have assumed a QSO $n(z)$ that is too wide for their actual $n(z)$. In addition, the samples of \cite{Petter2023} show a steeper correlation function  at small scales which  may  indicate a  contribution from a 1-halo term than if the correlation function followed a pure power-law. So to  understand the form of the ATLAS QSO correlation function in more detail, including the relative contributions of the 1- and 2-halo terms, in Section~\ref{sec:HOD_model_intro} we shall fit more sophisticated HOD models to our results.

\subsection{QSO bias and halo mass via $w_{qq}$}
\label{sec:wqq_bias}

We now compare our measurement of $w_{qq}$ (deprojected via Limber's formula) to the $\Lambda$CDM matter clustering correlation function at $z\approx1.7$ to estimate the QSO bias at this redshift. Following \cite{Croom2005}, we make this comparison by integrating over $\xi(r)$ out to r=20h$^{-1}$Mpc to form $\xi_{20}$. This $0<r<20$h$^{-1}$Mpc range is chosen so that it is dominated by the linear regime at $r>5$h$^{-1}$Mpc and where $\xi(r)$ can be approximated by a power-law:

\begin{equation}
\begin{aligned}
     \xi_{20}=3/20^3\int^{20}_0{\xi(r)r^2dr}=3/20^3\int^{20}_0{(r/r_0)^{-\gamma}\,r^2dr} \\
     \label{eq:xi_20}
\end{aligned}
\end{equation} 

\noindent Here we have already assumed a power law form for $\xi(r)$ with power-law slope, $-\gamma$, and scale length, $r_0$. We note that  $r_{\rm comoving}=20 {\rm h}^{-1}$Mpc  at $z=1.7$ corresponds to $\theta=20.'8$ (see Figs. \ref{fig:wqq} and \ref{fig:wqq_HOD}). \\

\begin{figure}
    \centering
	\includegraphics[width=\columnwidth]{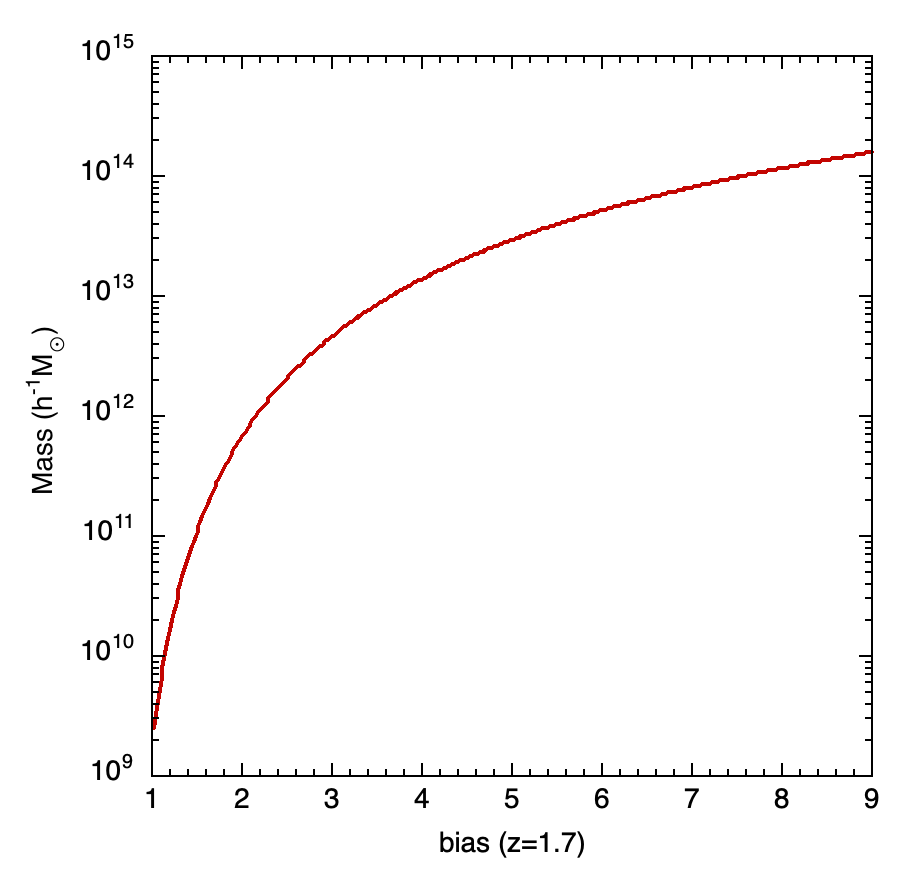} 
	\caption[Bias-mass relation for QSOs at redshift z=1.7]{The calculated bias-mass relation for QSOs at redshift z=1.7 based on equations 13-17 of \cite{Chehade2016}.}
	\label{fig:bias_mass_relation}
\end{figure}

Now, approximating $w_{qq}$ by a power-law of slope $1-\gamma=-0.67$ and applying Limber's formula, we find $r_0=5.3\pm{0.1}$h$^{-1}$Mpc  which from eq (\ref{eq:xi_20}) gives $\xi_{20}=0.25\pm{0.01}$ for our QSOs. A $\Lambda$CDM matter power spectrum implies $\xi_{20}=0.235$  for the matter at $z=0$. Assuming a linear gravitational growth factor of  $D(z=1.7)=2.033$  between $z=1.7$ and $z=0$  then gives $\xi_{20}=0.235/2.033^2=0.057$ for the matter at $z=1.7$. The QSO bias at $z=1.7$ is $b_Q=\sqrt(0.25/0.057)=2.09\pm{0.09}$. Then, following eqs. (13-17) of \cite{Chehade2016} we derive the bias-mass relation for QSOs at our average QSO redshift, $z=1.7$, as shown in Fig.~\ref{fig:bias_mass_relation}. From that, we estimate a mean QSO halo mass of $M_{\rm halo}=8.5\pm3\times10^{11}$ h$^{-1}$ M$_\odot$ at this redshift. 


\section{QSO-CMB Lensing Cross-Correlation}
\label{sec:G19cmb}

The cross correlation of our QSO catalogue with the Planck CMB lensing convergence map \citep{Planck2018} is an independent method towards determining the QSO bias and the QSO halo mass.

\subsection{Quasar-CMB Lensing Cross-Correlation Model}
\label{cmblensing_model}

We first perform our analysis using the model described in the studies of \cite{GeachPeacock2017}, \cite{Geach2019}, and \cite{Petter2022}. The model includes a lensing convergence contribution made by a $1-$halo and a $2-$halo term. Similarly to the equation used in the \cite{WI1998} model from \citetalias{EltvedtPaper2}, the convergence due to the 1-halo term is defined as:

\begin{equation}
     \kappa_1(R)=\frac{\Sigma(R)}{\Sigma_{crit}} 
     \label{eq:onehaloterm}
\end{equation} \\

\noindent where $\Sigma(R)$ is the projected mass surface density given an NFW density profile, and $\Sigma_{crit}$ is the critical surface density. Here, the projected mass surface density is: 

\begin{equation}
     \Sigma(R)=2\int^{\infty}_{R}\frac{r\rho(r)}{\sqrt{(r^2-R^2)}}dr 
     \label{eq:projecteddensity}
\end{equation} \\

The $2-$halo term is described by:

\begin{equation}
     \kappa_2(\theta)=\frac{\Bar{\rho}(z)}{(1+z)^3\Sigma_{crit}D^2(z)}\int\frac{ldl}{2\pi}J_0(l\theta)b_h\Delta (k,z)
     \label{eq:twohaloterm}
\end{equation} \\

\noindent where $J_0$ is a Bessel function of the zeroth order, $D(z)$ is the angular diameter distance, $\Delta(k,z)$ is the linear matter power spectrum, $\Bar{\rho}(z)$ is the average density of the Universe at $z$ and $b_h$ is the quasar bias for a halo of mass $M_h$. Then, the final model for the lensing convergence is:

\begin{equation}
     <\kappa>=\int dz (\kappa_1+\kappa_2)dn/dz
     \label{eq:combinedhaloterm}
\end{equation} \\

The lensing convergence results obtained by \citet{Geach2019} are shown in their Figure 3. There we see the radial profile of the quasar stacked convergence along with the best fitting lensing model as a solid line. The model includes the 1- and 2-halo contribution to the lensing signal, although at scales of $\theta>5'$, the $1-$halo term is mostly filtered out. A similar approach is taken by \cite{Petter2022}, where their Figure 5 shows a comparable model.  Both \cite{Geach2019} and \cite{Petter2022} filter and stack their model (based on Equation~\ref{eq:combinedhaloterm}) in order to mimic the filtering done on the CMB and QSO data. Therefore, the final model does fall below zero at $\sim\theta=40$arcmin (see Fig.~\ref{fig:QSO_cmb_xcorr_comparisons}), even though neither the 1- or 2-halo term components do so.  Below, we shall be simply scaling the 2-halo term of the model presented in both \cite{Geach2019} and \cite{Petter2022} for a first order calculation of our QSO bias. \\

The CMB lensing convergence denoted by $\kappa$ is a projection of a 3D density field. The quasar density is also a projection of a 3D density field. We convert these 3D projections into angular comoving distances in order to perform angular correlations. In our analysis we assume that we have the same absolute magnitude range as \cite{Chehade2016}. We also use a comparable quasar sample to \cite{Geach2019} and \cite{Petter2022}. However, our photometric redshifts are less accurate and therefore it may not be worth splitting into redshift and/or magnitude bins to perform further analysis.
 
\subsection{Quasar-CMB Lensing Cross-Correlation Results}
\label{sec:qsocmb_results}



\begin{figure}
    \centering
	\includegraphics[width=\columnwidth]{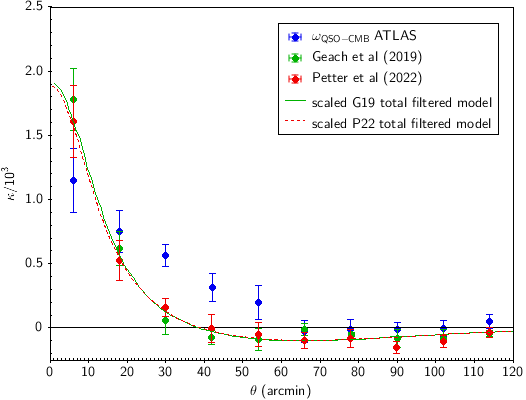} 
	\caption{We show our QSO-CMB Lensing cross-correlation result in blue along with the result obtained by \protect\cite{Geach2019} and \protect\cite{Petter2022}, in green and red respectively. The total filtered model, scaled by a factor of 0.8, from \protect\cite{Geach2019} is shown as a green line. The total filtered model, scaled by a factor of 0.85, from \protect\cite{Petter2022} is shown as a dashed red line. }
	\label{fig:QSO_cmb_xcorr_comparisons}
\end{figure}

Results of the cross-correlation we perform between our quasar sample and the Planck CMB Lensing map can be seen in Fig.~\ref{fig:QSO_cmb_xcorr_comparisons}. We show our results along with the results found by \citet{Geach2019} and \cite{Petter2022}. We note that the errors on our results are reasonably comparable to both these authors. The main difference between the two results is between $30-60'$ where our results are higher.  So, while the models of \cite{Geach2019} and \cite{Petter2022} fit their data at $\theta<60'$ with values of reduced $\chi^2\approx4\times10^{-4}$, (with only appeal left to a statistical fluctuation to explain this small $\chi^2$ value, given the consistency between their errors and ours),  these (scaled) models are rejected as fits to our result, both giving  reduced $\chi^2\approx8.4$ over the same $\theta<60'$ range. Comparing the observed results of ourselves and \cite{Geach2019} directly and so taking their errors into account, we still find a  reduced $\chi^2=3.2$ over the full $\theta<90'$ range. The reason for this discrepancy is unclear.


We first determine the QSO bias found with our results by scaling the total filtered  model determined by \cite{Geach2019}, indicated on Figure 3 of their paper, as well as the filtered total model from \cite{Petter2022}, indicated on Figure 5 of their paper, to our data. Both models include the 1- and 2-halo term described in the previous section and are filtered in the same way as their data. We look at the $0'<\theta<60'$ range as our data falls below $\kappa=0$ at larger scales. The reason for the  negative cross-correlations seen  at $40<\theta<60'$ by the other authors but not by ourselves is unclear. The main procedural difference was that  we did not make a tangential plane projection to the lensing data before stacking, so that may be the cause. Additionally, there are most likely more systematics at larger scales, despite the errors being smaller. We also note that this $\chi^2$ fit and associated errors are only approximate as they do not take into account covariance between data points. 

Upon scaling the models, we find that the total model from \cite{Geach2019} has a best $\chi^2$ fit for a scale of 0.8. Therefore, scaling the measured quasar halo bias of $b_h=2.7\pm0.3$ at $z=1.7$ found by \cite{Geach2019} by 0.8, gives us a QSO halo bias of $b_h=2.16\pm0.43$ at $z=1.7$. For the total model of \cite{Petter2022}, we find a best fit scaling factor of 0.85. Through scaling their QSO halo bias of $b_h=2.35\pm0.02$ by 0.85 we can infer a QSO bias value of $b_h=2.0\pm0.17$ for our data. The bias to mass relation described in Section~\ref{sec:wqq_bias} then indicates a host halo mass of $9.17$x$10^{11}h^{-1}$M$_{\odot}$ for $b_h=2.16$ and $6.71$x$10^{11}h^{-1}$M$_{\odot}$ for $b_h=2.0$. We therefore average these two bias measurements determined via scaling the total models of \cite{Geach2019} and \cite{Petter2022} to get $b_h=2.08\pm0.3$, with a host halo mass of $8.3\pm3\times10^{11}h^{-1}$M$_{\odot}$.

\begin{figure}
    \centering
	\includegraphics[width=\columnwidth]{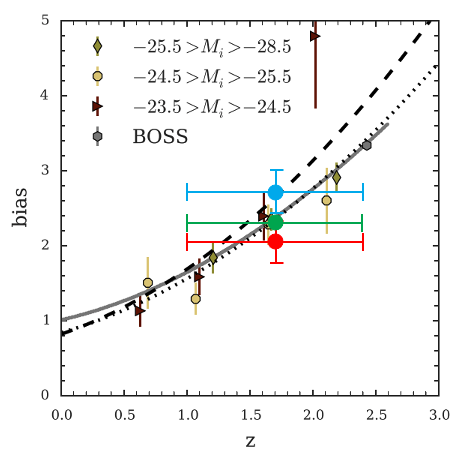} 
	\caption[The quasar halo bias as a function of redshift, taken from Figure 14 of \protect\cite{Chehade2016}, with our values of $b_h$ and the value determined by \protect\cite{Geach2019} and \protect\cite{Petter2022}.]{The quasar halo bias as a function of redshift, taken from Figure 14 of \cite{Chehade2016}. We show the $b_h$ value of $2.7\pm0.3$ at $z=1.7$ found by \cite{Geach2019} in blue and the $b_h=2.35\pm0.02$ value found by \cite{Petter2022} in green. Our bias value of $b_h=2.08\pm0.3$ estimated via scaling of the 2-halo model is shown together with our bias value of $b_h=2.09$ determined via $\omega_{qq}$ in Section~\ref{sec:wqq_bias} in red. The dotted black line represents the bias result determined by \cite{Chehade2016} and the bias result from 2QZ \citep{Croom2005} is shown as a dashed black line. The solid grey line represents the evolution for a halo of mass $2\times10^{12}$h$^{-1}$M$_{\odot}$.}
	\label{fig:C16_biases}
\end{figure}

In Figure 14 of \cite{Chehade2016}, they show the bias they determined as a function of redshift and absolute magnitude. In Fig.~\ref{fig:C16_biases}, we add to Fig. 14 of \cite{Chehade2016} our bias measurement of $b_h=2.08\pm0.3$, shown as a red point, and the $b_h=2.35\pm0.02$ value found by \cite{Petter2022} in green. The bias result of of $b_h=2.7\pm0.3$ at $z=1.7$ found by \cite{Geach2019} is shown as a blue point. We also show the bias value of $b_h=2.09$ determined via the QSO auto-correlation in Section~\ref{sec:wqq_bias} in yellow. The dotted black line represents the bias result determined by \cite{Chehade2016} and the bias result from 2QZ \citep{Croom2005} is shown as a dashed black line. Also in the figure is the measurement of the quasar halo bias from the BOSS survey \citep{BOSS2013} determined by \cite{Eftekharzadeh2015}. The solid grey line represents the evolution for a halo of mass $2\times10^{12}$h$^{-1}$M$_{\odot}$. From this figure, we see that the quasar halo bias measured by \cite{Geach2019} is in line with the bias measured 2QZ, but falls above the bias found by \cite{Chehade2016} (the black dotted line). The bias found by \cite{Petter2022} is in line with the result found by \cite{Chehade2016}. We see that our bias measurements fall slightly below all of these results but are still within reasonable agreement. We note that in applying this scaling process, we have effectively assumed the same redshift distribution, $dn/dz$, for ourselves, \cite{Petter2022}, and \cite{Geach2019}. Uncertainties in photometric redshifts and the ranges used for the analyses may thus also account for some of the  discrepancies between these results.


Overall there seems to be good agreement between the quasar-CMB lensing results of \cite{Geach2019}, \cite{Petter2022}, and the results from \cite{Chehade2016} which are derived from QSO clustering. The bias, and associated host halo mass results we find via QSO-CMB lensing are also in good agreement when scaling to the total model. We use this as a first order estimate of our data and continue forward by fitting a separate HOD model to our results. These QSO bias and host halo mass measurements, along with the measurements found in Section~\ref{sec:HOD_model_intro}, are summarized in Table~\ref{tab:QSO_biasandmass}.

\begin{table}\centering
 \caption[Summary of results for QSO bias $b_h$ and halo mass $M_{halo}$. Rows 1-3 contain our VST ATLAS results via (1) the  bias from the QSO angular autocorrelation function, (2) the bias fitted to the QSO-CMB lensing cross-correlation function and (3) a HOD model fitted jointly to the above two.]{Summary of results for QSO bias $b_Q$ and halo mass $M_{halo}$. Rows 1-3 contain our VST ATLAS results via (1) the  bias from the QSO angular autocorrelation function, (2) the bias fitted to the QSO-CMB lensing cross-correlation function and (3) a HOD model fitted jointly to the above two. Also shown are previous results from \cite{Croom2005}, \cite{Chehade2016}, \cite{Geach2019}, \cite{Petter2022} and  \cite{Petter2023}.}
 \label{tab:example}
 \begin{tabular}{lccc}
  \hline
  $z=1.7$ QSO Sample & bias & Mass & log Mass\\
+ Method     & ($b_h$)& $(10^{12}h^{-1}M_{\odot}$)&  ($h^{-1}M_{\odot}$)\\
  \hline
  ATLAS ACF $w_{qq}$                     & $2.09\pm{0.09}$       & $0.85\pm0.3$              &   $11.9\pm0.04$ \\
  ATLAS CMB Lensing  $w_{q\kappa}$       & $2.08\pm0.3$          & $0.83^{+0.8}_{-0.5}$      &   $11.9\pm0.4$ \\
  ATLAS HOD $w_{qq}$+$w_{q\kappa}$       & $2.60^{+0.30}_{-0.23}$& $2.5^{+1.5}_{-0.9}$       &   $12.4\pm0.2$ \\ 
  Croom05 z-space ACF                    & $2.17\pm0.09$         & $1.2\pm0.2$               &   $12.1\pm0.08$ \\  
  Chehade16 z-space ACF                  & $2.34\pm0.35$         & $2.0\pm1.0$               &   $12.3\pm0.4$ \\  
  Geach19 CMB Lensing                    & $2.7\pm0.3$           & $4.0^{+2.3}_{-1.5}$       &   $12.6\pm0.2$ \\ 
  Petter22 CMB Lensing                   & $2.35\pm0.02$         & $3.0\pm0.4$               &   $12.5\pm0.05$ \\ 
  Petter23 HOD $w_{qq}$+$w_{q\kappa}$    & $2.3\pm0.5$           & $4.0\pm1.4$               &   $12.6\pm0.2$ \\ 
  \hline
 \end{tabular}
 \label{tab:QSO_biasandmass}
\end{table}

\section{HOD Model via QSO auto-correlation and QSO-CMB lensing cross-correlation}
\label{sec:HOD_model_intro}

\subsection{HOD model}
\label{sec:HOD_model}

We utilize the CHOMP package based on the work of \cite{Jain2003} and \cite{Scranton2005} and available on the CHOMP GitHub (CHOMP; \href{https://github.com/karenyyng/chomp/blob/master/README.txt}{Morrison, Scranton, and Schneider}), to fit HOD models to both the auto-correlation, $w_{qq}$, of our QSO sample as well as the QSO-CMB lensing convergence map cross-correlation, $w_{g\kappa}$. To check the CHOMP methodology, we first supplied CHOMP with the HOD model parameters of \cite{Petter2023}. However, in the case of the QSO correlation function, we found that CHOMP could not reproduce the $w_{qq}$ results of \cite{Petter2023}, assuming  their HOD parameters. In this case, we used the alternative HaloMod package \citep{HaloMod2021} to predict the \cite{Petter2023} 3-D $\xi(r)$ and then input this into Limber's formula using Eq.(13) of \cite{Phillipps1978}. The resulting $w_{qq}$ was found to agree with \cite{Petter2023} and the same procedure was then used to fit our ATLAS $w_{qq}$.

For the 1-halo term of QSO and matter clustering, we assume a \cite{NFW1996} NFW model, to predict the projected, lensed mass profile, $w_{qm}$ and then $w_{q\kappa}$. 
CHOMP also assumes that halo concentration is a function of halo mass with the functional form  $c(m)\approx9(m/m^*)^{-0.13}$ taken  from \cite{Bullock2001}. For the 2-halo term, CHOMP assumes the form given in eq. (6) of \cite{Jain2003} with a bias model from \cite{Tinker2010}, etc. CHOMP and HaloMod both allow use of the 5-parameter HOD model of \cite{Zheng2007} to fit the 2-point auto-correlation function. For all  HOD models, we assume a $\Lambda$CDM cosmological model with the matter density $\Omega_{M}=0.3-0.046$, baryon density $\Omega_{b}=0.046$ and $\Omega_{\Lambda}=0.7$. We assume adiabatic Gaussian primordial density fluctuations with a power-law index of the spectrum $n_s=0.96$. The r.m.s. matter density fluctuation is assumed to be $\sigma_8=0.8$. The Hubble constant we use throughout is $h=0.7$. Finally, we define the QSO halo redshift to be at $z=1.7$ and the CMB at redshift $z=1100$. Then we fit the  HOD parameters: the minimum halo mass scale, $\log M_{min}$, the minimum mass scale softening width, $\sigma_{\log M}$, for the  central galaxies, the satellite cut-off  mass scale, $\log M_0$, the satellite HOD power-law normalization, $\log M_{1'}$ and its slope,  $\alpha$, at high halo masses.

\subsection{QSO autocorrelation}

\begin{figure}
    \centering
 	\includegraphics[width=\columnwidth]{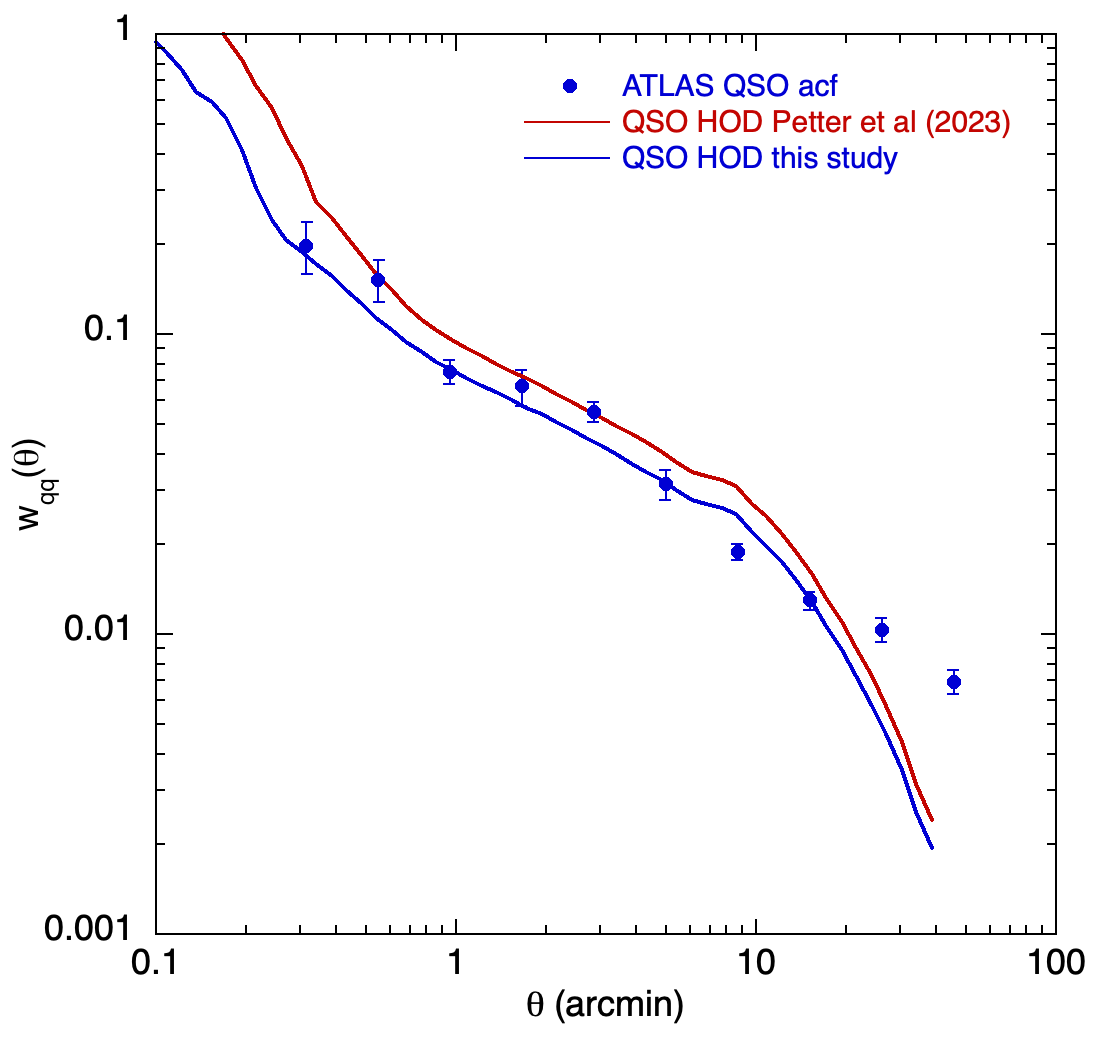} 
	\caption{The ATLAS $17<g<22$ QSO angular auto-correlation function, $w_{qq}$, compared to our $1<z<2.2$ HOD model (solid blue line) with parameters: $\log (M_{min})=12.2$, $\sigma_{\log M}=0.4$, $\log (M_0)=12.2$, $\log(M_{1'})=13.28$ and $\alpha=0.7$. The red line shows the unobscured QSO HOD model of \protect\cite{Petter2023} with parameters: $\log (M_{min})=12.42$, $\sigma_M=0.4$, $\log (M_0)=12.42$, $\log(M_{1'})=13.28$ and $\alpha=0.7$. Both HOD models  assume standard $\Lambda$CDM parameters with $h=0.7$ and $\sigma_8=0.8$.}
	\label{fig:wqq_HOD}
\end{figure}

In Fig.~\ref{fig:wqq_HOD}, we compare our observed QSO correlation function, $w_{qq}$, to our fitted HOD (solid blue line), with parameters, $\log M_{min}=\log M_0=12.2$, $\log M_1=13.28$, $\sigma_{\log M}=0.4$ and $\alpha=0.7$. Also shown as the solid red line is the HOD fit from \cite{Petter2023} for their unobscured  QSO sample, with parameters $\log M_{min}=\log M_0=12.4$, $\log M_1=13.5$, $\sigma_{\log M}=0.4$ and $\alpha=0.7$. This model is rejected by our data to $\theta<15'$ with a reduced $\chi^2$ of 7.2. We see that our HOD model has slightly reduced mass parameters due to the ATLAS $w_{qq}$ having a lower amplitude ($r_0=5.2$ h$^{-1}$ Mpc) than the unobscured QSO $w_{qq}$ of \cite{Petter2023} that has $r_0=6.0$ h$^{-1}$ Mpc, assuming a $\gamma=-1.8$ power-law form for $\xi(r)$ in both cases (as noted in Sec.~\ref{sec:Limber_approx}). Our HOD fits the data to $\theta<15'$ with a reduced $\chi^2$ of 1.79.

\subsection{QSO-CMB cross-correlation}

We next test these two HOD models for internal and external consistency using the QSO-CMB lensing convergence map cross-correlation function. As we see some discrepancy with regards to various parameter fits for the QSO autocorrelation, we use the QSO-CMB cross-correlation as an independent method to measure the host halo masses. Here, we expand upon the results described in Section~\ref{sec:qsocmb_results} to compare the predictions of the above two  HOD models to our QSO-CMB cross correlation results.  We note that at the $6'$ resolution of the Planck lensing map, our cross-correlation analysis will mainly be sensitive to the 2-halo term of QSO clustering.

\begin{figure}
  \centering
  \includegraphics[width=\columnwidth]{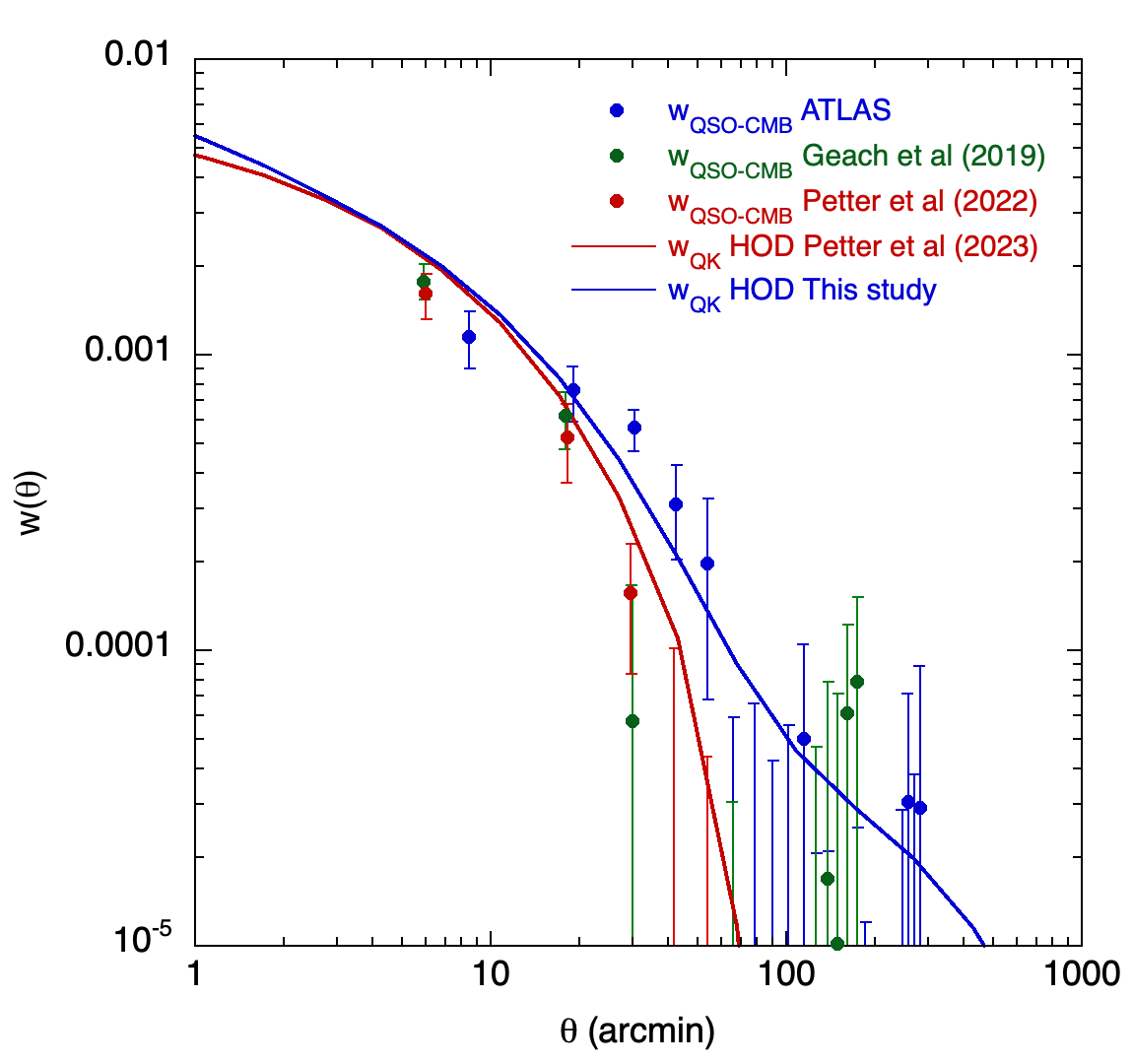}
  \caption{The ATLAS QSO-Planck CMB Lensing Map angular cross-correlation function, $w_{QSO-CMB}$, compared to the results of  \protect\cite{Geach2019} and \protect\cite{Petter2022}. The solid blue line represents our $1<z<2.2$ QSO HOD model previously shown  in Fig. \ref{fig:wqq_HOD} with the same HOD parameters as detailed there. The solid red model represents the HOD model of \protect\cite{Petter2023} again as shown in  Fig. \ref{fig:wqq_HOD}. Both models again assume standard $\Lambda$CDM parameters with  $h=0.7$ and $\sigma_8=0.8$.}
  \label{fig:wq_cmb_hod}
\end{figure}

We see in Fig. \ref{fig:wq_cmb_hod} that our observed  CMB lensing cross-correlation function is in good agreement with our HOD model,  with the model giving a value of reduced $\chi^2=1.04$ when fitted over the full $\theta$ range shown. As previously noted in Section \ref{sec:qsocmb_results} our observational results are only in approximate agreement with the results of \cite{Geach2019} and \cite{Petter2022}, since our cross-correlation data points are significantly higher than the other two at $10<\theta<60'$. This increased amplitude of our results then appears to be reflected in the HODs derived from the ATLAS and \cite{Petter2023} $w_{qq}$ results, as discussed in Section \ref{sec:halo_mass} below.



\begin{figure}
    \centering
 	\includegraphics[width=\columnwidth]{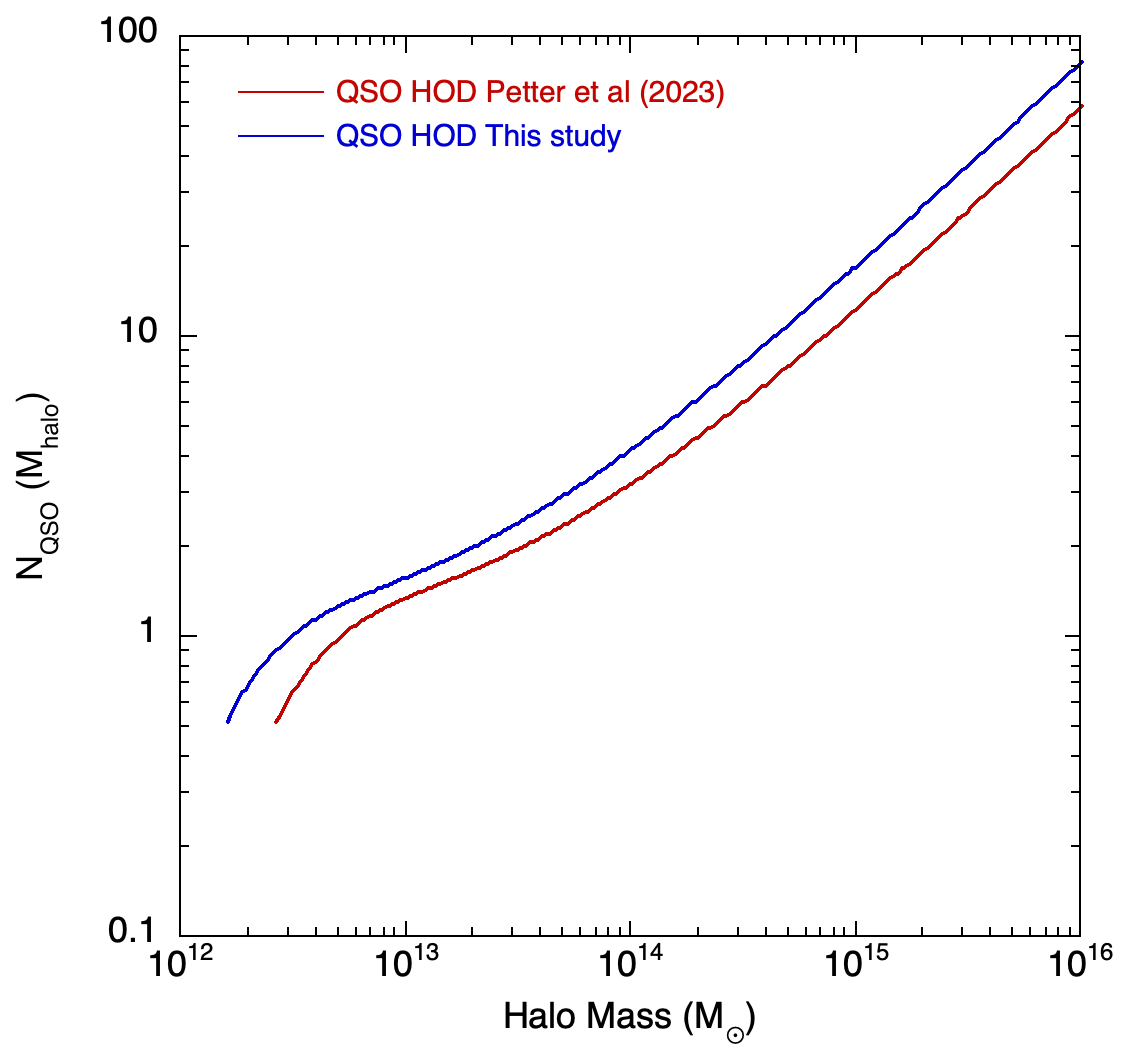}
	\includegraphics[width=\columnwidth]{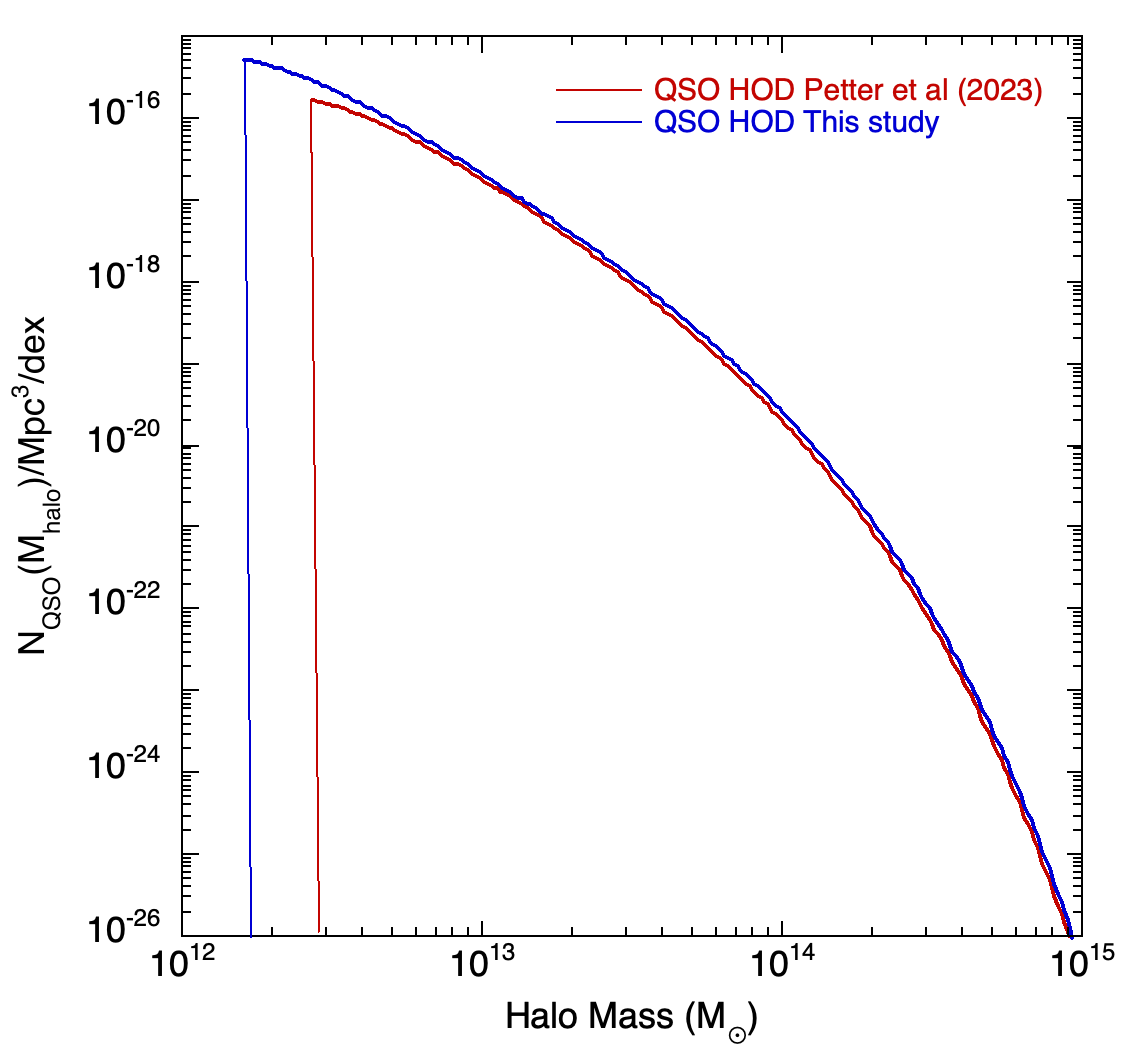}
  \caption{(a) The QSO HOD as a function of halo mass. The solid blue line represents our   $1<z<2.2$ QSO HOD model previously shown  in Fig. \ref{fig:wqq_HOD} with the same HOD parameters as detailed there. The solid red model represents the HOD model of \protect\cite{Petter2023} again as shown in  Fig. \ref{fig:wqq_HOD}. (b) The QSO space density as a function of halo mass formed by multiplying the halo mass function by a HOD model. The solid blue line represents our   $1<z<2.2$ QSO HOD model previously shown  in Fig. \ref{fig:wqq_HOD} with the same HOD parameters as detailed there. The solid red model represents the HOD model of \protect\cite{Petter2023} again as shown in  Fig. \ref{fig:wqq_HOD}.} 
	\label{fig:Nqso_hod}
\end{figure}

\subsection{Halo mass of QSOs}
\label{sec:halo_mass}

In Fig. \ref{fig:Nqso_hod}(a) we show our QSO HOD in blue, using the same parameters as in Fig.~\ref{fig:wqq_HOD}, alongside the \cite{Petter2023} HOD in red, as a function of halo mass, i.e. the halo mass function. In Fig. \ref{fig:Nqso_hod}(b) we show the QSO space density as a function of halo mass formed by multiplying the halo mass function by our HOD model and that of \cite{Petter2023}. Averaging over these two distributions we find that the average QSO halo mass is $\log M_{eff}=12.39$ ($h^{-1}M_\odot$) for our HOD and $\log M_{eff}=12.90$ ($h^{-1}M_\odot$) for the HOD of \cite{Petter2023}. We note that this latter value calculated by ourselves is somewhat higher than the $\log M_{eff}=12.6 ( h^{-1}M_\odot$) found by \cite{Petter2023} based on their $b_{eff}=2.25$. Otherwise we note that our lower ($\approx3\times$) average QSO halo masses are in line with our $w_{qq}$ having a lower scale length of $r_0=5.2 $ h$^{-1}$ Mpc compared to the $r_0=6.0$ h$^{-1}$ Mpc measured by \cite{Petter2023} for their unobscured QSO sample. Also the results we obtain from the HOD analyses are generally slightly larger than when estimating halo masses directly from $w_{qq}$ and $b_Q$. However, in terms of the QSO-CMB lensing results in Fig. \ref{fig:wq_cmb_hod}, we do find a larger observed  amplitude at $\theta>10'$ than \cite{Geach2019} for this cross-correlation which is well fitted by our HOD model. The lower cross-correlation at $\theta>10'$ observed by \cite{Petter2023} is also well fitted by their HOD model. So the HOD models seem internally consistent between $w_{qq}$ and $w_{QSO-CMB}$. The HODs also seem externally consistent, given the similarities in halo mass function and average halo masses between \cite{Petter2023} and ourselves.


Working directly from our QSO halo mass function in Fig. \ref{fig:Nqso_hod} (b), we find that the average halo mass at $z=1.7$ is $\log M_{eff}=12.4$. We further note that our $\log M_{eff}=12.4$ QSO HOD estimate for the average QSO halo mass is higher than the $\log M=11.9$ QSO halo mass found from the $b_Q=2.09$ bias implied by the ratio of the QSO and matter auto-correlation functions (see Section \ref{sec:wqq_bias}). It is also larger than the halo mass estimated at $M=2.58\pm^{0.39}_{0.36}\times10^{12}$h$^{-1}$M$_{\odot}$ i.e. $\log M=11.9$ from the bias of $b_Q=2.08\pm0.3$ we found by scaling the QSO CMB lensing result of \cite{Geach2019} in Section \ref{sec:qsocmb_results}, shown in Table~\ref{tab:QSO_biasandmass}.

From our HOD model in Fig. \ref{fig:Nqso_hod} (b) we see that the QSO mass function shows a steep fall away from $\log M_{min}=12.2$. Indeed, we calculate that 67\% of these $z=1.7$ QSOs lie in the small halo mass range $12.2<\log M <13.2$. Thus it is not unreasonable to say that most QSOs have roughly the same halo mass. This result has been noted before by \cite{Shanks2011} who compared the increasingly faint SDSS, 2QZ and 2SLAQ measurements of the redshift-space correlation $\xi(s)$  and found no dependence of the QSO clustering amplitude on QSO luminosity. \cite{Chehade2016} in the 2QDES pilot survey reached even fainter magnitudes and confirmed this luminosity independence over an order of magnitude (i.e. $\approx10\times$) in luminosity and over the full $0<z<2.5$ redshift range. Since QSO clustering is luminosity independent the implication may be that QSO halo mass and then QSO black hole mass are also luminosity and redshift independent. Here our HOD model now shows that despite the large luminosity range shown by QSOs  at fixed redshift (e.g. $z\approx1.7$), the range of halo and hence black hole masses covered is actually very small. This supports the idea that most QSOs have the same black hole mass.

\subsection{QSO halo mass and stellar mass functions compared}
\label{sec:stellar_mass}

Further supporting evidence for this hypothesis comes from the X-ray survey in the PRIMUS field, analysed by \cite{Aird2012}. Their Figure 4 (top left panel)  shows the probability, $p(L_X | M_*,z)$ for a galaxy of given stellar mass, $M_*$, and redshift, $z$, to host an AGN of X-ray luminosity, $L_X$ found for X-ray emitting AGN in this survey. Although it is clear that there do exist X-ray AGN in low stellar mass galaxies, their numbers are quite small compared to the numbers in high stellar mass galaxies.

\begin{figure}
    \centering
 	\includegraphics[width=\columnwidth]{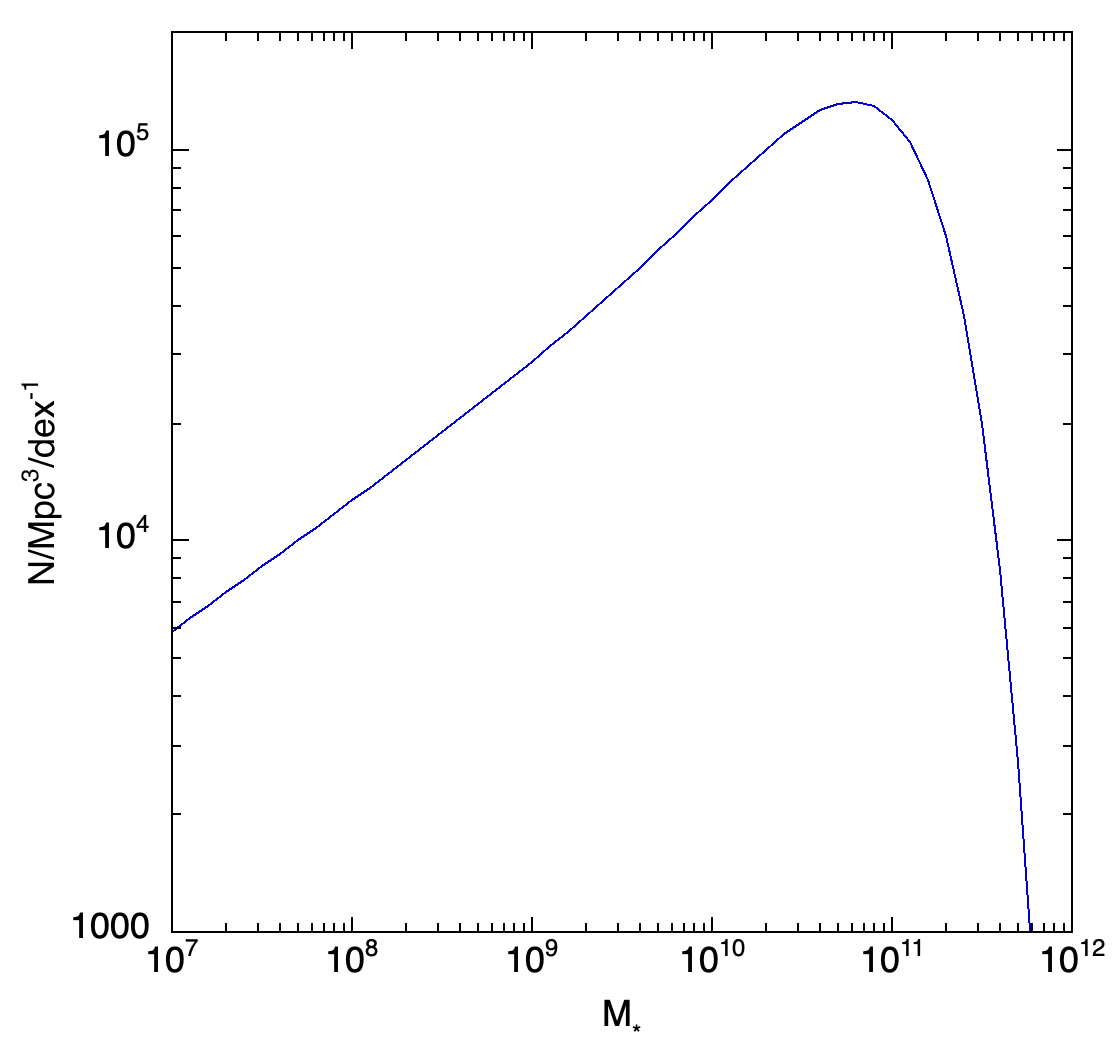}
  \caption{The QSO space density as a function of galaxy stellar mass formed by multiplying the galaxy stellar mass function by essentially the X-ray luminosity-stellar mass relation of \protect\cite{Aird2012}. Note the similarity to the QSO space density as a function of halo mass in Fig. \ref{fig:Nqso_hod}.}
	\label{fig:Nqso_aird}
\end{figure}

From the top left panel of their Fig. 4 with $0.2<z<0.6$, we sum over the four $L_X$ bins to find the probability, $p_{AGN}(M^*,z)$, of a galaxy of stellar mass, $M_*$, hosting an X-ray AGN. Since the relations in the four $L_X$ bins appear approximately parallel, we are justified here in adopting an average slope giving $p_{AGN}\approx M_*^{0.75}$. From \cite{Ilbert2013}, their 0$.2<z<0.5$ stellar mass function in their Fig. 5 is given by their eq (2)  with $\log(M*)=10.88$, $\phi*_1=1.68\times10^{-3}$ h$^3$Mpc$^-3$, $\alpha_1=-0.69$, $\phi*_2=0.77\times10^{-3}$ and $\alpha_2=-1.42$. Multiplying this stellar mass function by the probability,  $p_{AGN}\sim M_*^{0.75}$, then gives the number density of AGN as a function of stellar mass as shown in Fig. \ref{fig:Nqso_aird}. So for the form of the AGN-stellar mass function we find a peaked distribution, centred on $M_*\approx6\times10^{10} M_\odot$. Essentially the low stellar mass end is cut off by the steep correlation with X-ray luminosity while the AGN space density at high stellar masses is naturally suppressed by the decrease in the galaxy stellar mass function at high masses. We find 67\% of the AGN lie in the stellar mass range $10.0<\log M_*(M_{\odot})<11.5 $. Given a halo mass-stellar mass ratio of $\approx40$ at the peak we see that there is reasonable consistency between the QSO halo and stellar mass functions with both implying a relatively small mass range preferred for QSO hosts.

\begin{figure}
    \centering
  	\includegraphics[width=\columnwidth]{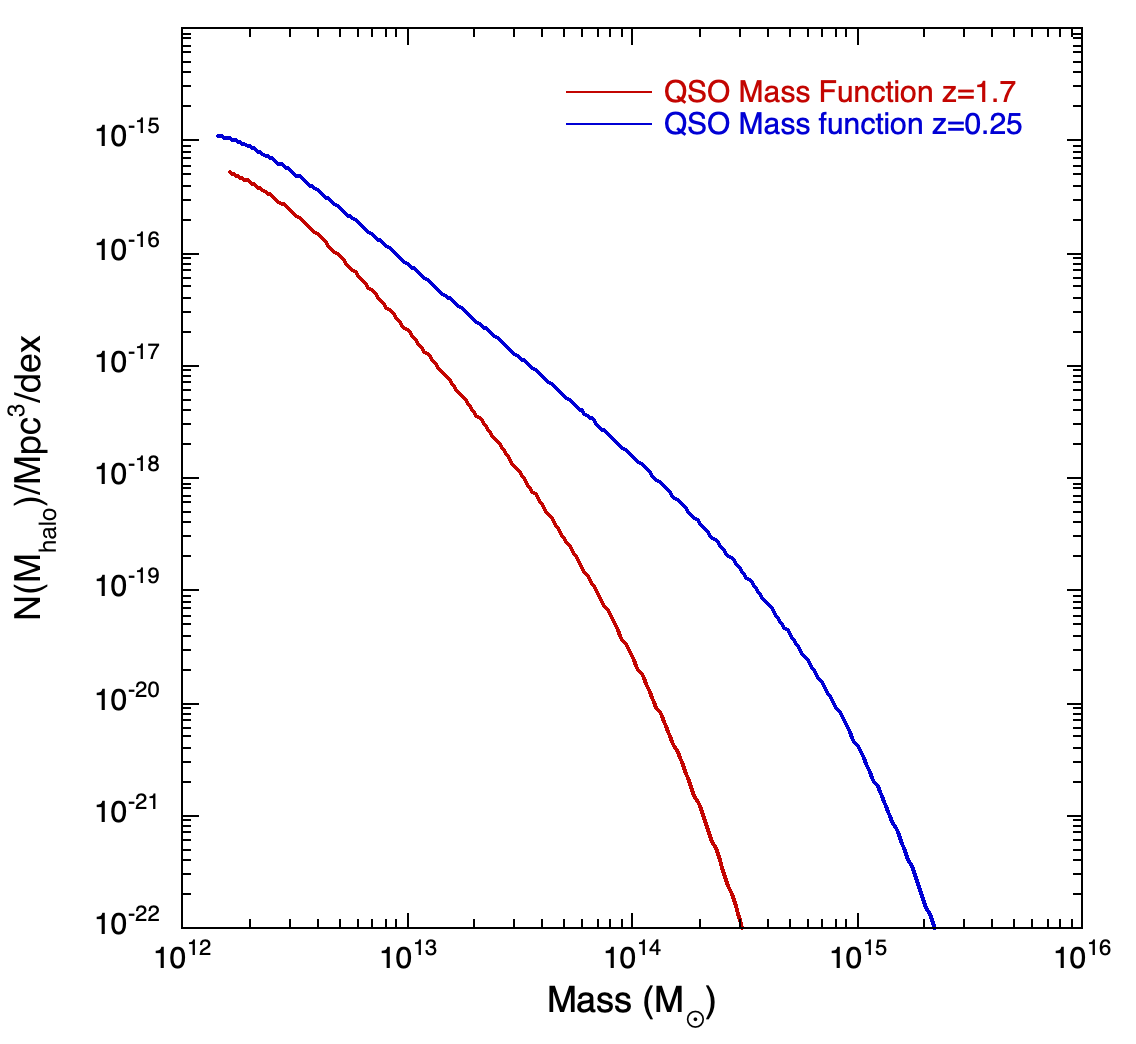}
      \caption[The QSO space density at $z=1.7$ and $z=0.25$ as a function of halo mass for our  QSO HOD model previously shown  in Fig. \ref{fig:wqq_HOD} and with the same HOD parameters as detailed there.]{The QSO space density at $z=1.7$ and $z=0.25$ as a function of halo mass for our  QSO HOD model previously shown  in Fig. \ref{fig:wqq_HOD} and with the same HOD parameters as detailed there. The difference between the two models represents the evolution of the halo mass function between these two redshifts which  may appear not dis-similar to the Pure Luminosity Evolution shown by the QSO Luminosity Function.}
	\label{fig:Nqso_hod_z}
\end{figure}

\subsection{Evolution  of QSO halo mass and luminosity functions}
\label{sec:LF}

Finally, we consider the question of the physical interpretation of the evolution of the QSO Luminosity Function (LF) which takes a Pure Luminosity Evolution (PLE) form over the range $0<z<2.2$ (e.g. \citealt{Longair1966, Marshall1985}). Here, QSO luminosity at $L^*$ increases by $\approx(1+z)^3$ or a factor of $\approx30$ by $z=2.2$. \cite{Shanks2011} (see also \citealt{Marshall1985, Boyle1988, Croom2004, Croom2009}) have speculated that this coherence of the amplitude of the QSO LF over $\approx10^{10}$ years might demand a coincidence if QSOs only had lifetimes of $10^6-10^9$ years, implying that they may be longer-lived. The PLE model also predicts only a slow rise in BH mass between $z=2.2$ and $z=0$ despite the sharp decrease in QSO luminosity and again this could be in agreement with the luminosity independence of QSO clustering. However, these authors also noted that the evolution of the QSO correlation function and its implied bias with redshift were not consistent with such a long-lived model (see e.g. \citealt{Fry1996}). The evolution of the QSO mass function shown in Fig. \ref{fig:Nqso_hod_z} highlights this issue. Here, taking $b_Q(z=0.25)=1.0$ from the $b_Q-z$ fit shown in eq. 15 of \cite{Croom2005}, and fitting a HOD to the resulting $\xi_{20}(z=0.25)$ with $\log (M_{min})=\log (M_0)=12.15$, $\sigma_{\log M}=0.4$, $\log(M_{1'})=13.23$ and $\alpha=0.7$ gives the QSO mass function for $z=0.25$. This has moved to higher masses compared to the $z=1.7$ case, taken from Fig. \ref{fig:Nqso_hod}(b), as might be expected after gravitational growth, whereas the LF moves to lower luminosities at lower redshift (see e.g. Fig. 6 of \citealt{Croom2009}). This  behaviour of the LF under PLE is sometimes called "downsizing". With lower luminosities and higher masses at low redshifts the Eddington ratios are clearly lower at low redshift. Now despite the opposite direction of the evolution of the mass and luminosity functions, both appear consistent with evolution in the horizontal (mass) direction with the mass function showing roughly constant space densities with redshift, similar to the LF. Since this constant QSO density with redshift seems to appear naturally out of gravitational growth in the $\Lambda$CDM model, then no appeal to a long-lived QSO model may be needed to explain the non-evolution of the QSO LF in the density direction. In this case the dimming of the QSO LF with decreasing redshift could be ascribed to  depletion of the supply of gas+stars to fuel  QSO accretion. The analogy here would be with galaxy evolution if the reason for the  observed dimming of the galaxy LF is depletion of the gas supply, in this case inhibiting  star-formation. Although in both cases there would need to be a mechanism invoked to restrict fuel supply while gravitational growth was on-going, this interpretation for QSO PLE seems worth further study, including testing its prediction of QSO $M_{BH}$ independence of redshift and luminosity via (stacked) reverberation mapping analyses.

\section{Conclusions}
\label{sec:conclusions}

The aim of this paper was to make new estimates of QSO host halo masses via QSO clustering  and  QSO-CMB lensing cross-correlation analyses. The QSO catalogues came from  the  VST-ATLAS quasar survey of \citetalias{Eltvedt2023}. The depth and reliability  of the ATLAS QSO catalogue meant that we could measure the QSO 2-point angular correlation function directly from the data. We found that it was well modelled  in 3-D by a correlation function with a power law form, $\xi(r)=(r/r_0)^\gamma$ with $r_0=5.2$h$^{-1}$ Mpc and $\gamma=-1.8$. Then assuming a linear regime mass power spectrum in a $\Lambda$CDM model, we compared galaxy and mass auto-correlation functions within a 20h$^{-1}$ Mpc radius sphere (i.e. $\xi_{20}$) to find $b_Q=2.09\pm0.09$ implying a QSO halo mass of $M_{halo}=8.5\pm0.3\times10^{11} M_{\odot}$.

We then cross-correlated the QSO sample described in Sec.~\ref{sec:xcorr_QSO_sample} with the CMB lensing maps of  \cite{Planck2018}. We first used methods similar to those outlined by \cite{Geach2019} to measure the bias and halo mass via the lensing of the CMB by foreground quasars. Here we find good agreement between our data and that of \cite{Geach2019} as well as that of \cite{Petter2022}. We are then able to fit the model determined by \cite{Geach2019} to our data with a scaling factor of $0.8$ and a scaling factor of $0.85$ to the model determined by \cite{Petter2022}. Therefore, we are able to measure a quasar halo bias of $b_h=2.08\pm0.3$ at an average redshift of $z=1.7$, corresponding to a halo mass of $0.83\times10^{12}h^{-1}$M$_{\odot}$. Our bias value is in excellent agreement with the quasar bias from quasar clustering in \cite{Chehade2016} as well as the QSO-CMB cross-correlation study of \cite{Petter2022}.


We then combined these two methods and fitted a HOD model to $w_{qq}$ which could be tested for consistency using QSO-CMB lensing. The HOD parameters that we obtained from $w_{qq}$ were similar to those measured for unobscured QSOs by \cite{Petter2023} with the only difference being that $\log (M_{min}$, $\log (M_0)$, and $\log (M_{1'})$ were $\log (M)=0.2$ smaller than measured by \cite{Petter2023}. From the resulting QSO mass function produced by multiplying the $\Lambda$CDM halo mass function by the QSO HOD, we found an average QSO halo mass at $z=1.7$ of $\log M_{eff}=12.4$, again about $\log (M)=0.2$ smaller than measured by \cite{Petter2023} and also slightly higher than measured from $b_h$ inferred directly from $w_{qq}$ and via CMB lensing cross-correlation, $w_{q\kappa}$. However, this HOD model from our $w_{qq}$ was also found to be a good fit to our CMB lensing results, confirming consistency between these two independent  observations. From the QSO mass function we also found that 67\% of the $z=1.7$ QSOs had halo masses that lie in the small halo mass range $12.2<\log M <13.15$ suggesting that most QSOs have similar halo and hence black hole masses. A similar result can be found by combining the galaxy stellar mass function of \cite{Ilbert2013} and the probability of a galaxy hosting an X-ray AGN as a function of stellar mass as estimated by \cite{Aird2012}. Here,  67\% of $z=1.7$ AGN are found to lie in the  stellar mass range $10.0<\log M_*(M_{\odot})<11.5$.

Finally, we inter-compared the QSO halo mass functions at $z=0.25$ and $z=1.7$ and showed that they appear to evolve to higher masses as redshift decreases as would be expected from gravitational growth. Although the QSO LF evolves in the opposite direction to lower luminosities at low redshift, the two functions otherwise appear similar with the halo mass function evolving mostly in the mass rather than the space density direction. This is reminiscent of the PLE shown by the QSO Luminosity Function. Thus it may be that this constant QSO space density with redshift may be naturally explained by gravitational growth in a $\Lambda$CDM Universe with no need to invoke long lived QSO models as discussed by \cite{Shanks2011}. In this case the decreasing brightness of QSOs towards the present day may be explained by the increasing lack of material available for accretion to fuel the QSO, despite the gravitational growth of the QSO halo mass. This would then amount to an analogous explanation to the PLE seen in galaxies where the low luminosity of galaxies at low redshift may be due to the reduction in the gas supply needed to fuel star-formation, causing the galaxies to dim in the rest optical bands by the present day (see e.g. \citealt{Metcalfe2001}).


\section*{Data Availability Statement}
The ESO VST ATLAS  data we have used are all publicly available.  The VST ATLAS QSO Catalogue can be found at \url{https://astro.dur.ac.uk/cea/vstatlas/qso_catalogue/}. All other data relevant to this publication will be supplied on request to the authors. 

\section*{Acknowledgements}

We acknowledge use of the ESO VLT Survey Telescope (VST) ATLAS. The ATLAS survey is based on data products from observations made with ESO Telescopes at the La Silla Paranal Observatory under program ID 177.A-3011(A,B,C,D,E.F,G,H,I,J,K,L,M,N) (see \citealt{Shanks2015}). BA acknowledges support from the Australian Research Council’s Discovery Projects scheme (DP200101068).
LFB acknowledges support from ANID BASAL project FB210003. We finally acknowledge STFC Consolidated Grant  ST/T000244/1 in supporting this research.

For the purpose of open access, the authors have applied a Creative Commons Attribution (CC BY) licence to any Author Accepted Manuscript version arising.

Finally, we thank the referee for useful comments that improved the quality of this paper.



\bibliographystyle{mnras}
\bibliography{bibliography} 








\bsp	
\label{lastpage}
\end{document}